\documentclass[a4paper,12pt]{article}
\pdfoutput=1
\usepackage{feynmp-auto,expdlist}
\usepackage{amsmath, amsfonts, amssymb}
\usepackage{graphicx}
\usepackage{enumerate}
\usepackage{hyperref}
\usepackage{latexsym}
\usepackage{dsfont}
\usepackage{hepnicenames}
\usepackage{enumerate}
\usepackage{soul}
\usepackage[normalem]{ulem}
\usepackage{comment}

\usepackage{units}

\usepackage{soul}

\usepackage{mathrsfs,graphicx,rotating,amsmath,amsfonts,mathtools,booktabs,amssymb,wasysym}
\usepackage{hyperref}\usepackage{slashed}
\usepackage[nosort]{cite}
\usepackage[table,xcdraw,dvipsnames]{xcolor}
\usepackage{bm}
\usepackage{graphicx}
\usepackage{multirow,multicol}
\hypersetup{colorlinks,bookmarksopen,bookmarksnumbered,
	linkcolor=blus,pdfstartview=FitH,urlcolor=rossos,citecolor=verde}
\allowdisplaybreaks

\renewcommand{\[}{\left[}
\renewcommand{\]}{\right]}
\renewcommand{\(}{\left(}
\renewcommand{\)}{\right)}

\def\Lag{\mathscr{L}}

\newcommand{\mio}[1]{}

\def\bpm{\begin{pmatrix}}
	\def\epm{\end{pmatrix}}

\usepackage{mathrsfs}

\newcommand{\fig}[1]{~\ref{fig:#1}}
\newcommand{\sfrac}[2]{#1/#2}

\allowdisplaybreaks
\usepackage{multicol}
\usepackage{color}
\definecolor{rosso}{cmyk}{0,1,1,0.4}
\definecolor{rossos}{cmyk}{0,1,1,0.55}
\definecolor{rossoc}{cmyk}{0,1,1,0.2}
\definecolor{blu}{cmyk}{1,1,0,0.3}
\definecolor{blus}{cmyk}{1,1,0,0.6}
\definecolor{bluc}{cmyk}{1,1,0,0.1}
\definecolor{verde}{cmyk}{0.92,0,0.59,0.25}
\definecolor{verdec}{cmyk}{0.92,0,0.59,0.15}
\definecolor{verdes}{cmyk}{0.92,0,0.59,0.4}

\newcommand\Ord{{\cal O}}
\newcommand{\eq}[1]{~{\rm (\ref{eq:#1})}}

\newcommand{\GeV}{\,{\rm GeV}}

\def\circa#1{\,\raise.3ex\hbox{$#1$\kern-.75em\lower1ex\hbox{$\sim$}}\,}

\newcommand{\beq}{\begin{equation}}
\newcommand{\eeq}{\end{equation}}

\newcommand{\bea}{\begin{eqnarray}}
\newcommand{\eea}{\end{eqnarray}}
\newcommand{\be}{\begin{equation}}
\newcommand{\ee}{\end{equation}}
\font\tenrsfs=rsfs10 at 12pt
\font\sevenrsfs=rsfs7
\font\fiversfs=rsfs5
\newfam\rsfsfam
\textfont\rsfsfam=\tenrsfs
\scriptfont\rsfsfam=\sevenrsfs
\scriptscriptfont\rsfsfam=\fiversfs

\newsavebox\MBox

\renewenvironment{thebibliography}[1]
{\begin{multicols}{2}[\section*{\refname}]%
		\@mkboth{\MakeUppercase\refname}{\MakeUppercase\refname}%
		\list{\@biblabel{\@arabic\c@enumiv}}%
		{\settowidth\labelwidth{\@biblabel{#1}}%
			\leftmargin\labelwidth
			\advance\leftmargin\labelsep
			\@openbib@code
			\usecounter{enumiv}%
			\let\p@enumiv\@empty
			\renewcommand\theenumiv{\@arabic\c@enumiv}}%
		\sloppy
		\clubpenalty4000
		\@clubpenalty \clubpenalty
		\widowpenalty4000%
		\sfcode`\.\@m}
	{\def\@noitemerr
		{\@latex@warning{Empty `thebibliography' environment}}%
		\endlist\end{multicols}}

\newcommand{\eV}{\,{\rm eV}}
\newcommand{\SU}{\,{\rm SU}}

\def\circa#1{\,\raise.3ex\hbox{$#1$\kern-.75em\lower1ex\hbox{$\sim$}}\,}
\makeatletter

\font\ital=cmu10

\def\hhref#1{\href{http://arxiv.org/abs/#1}{arXiv:#1}}
\usepackage{xstring}
\newcommand{\hhrefq}[1]{\IfSubStr{#1}{:}{\href{http://inspirehep.net/search?ln=en&ln=en&p=#1&of=hb&action_search=Search&sf=&so=d&rm=&rg=25&sc=0}{InSpire:#1}}{\hhref{#1}}}

\def\art{\@ifnextchar[{\eart}{\oart}}
\def\eart[#1]#2#3#4#5#6{{\rm #2}, {\em #3 \bf #4} {\rm (#6) #5} ({\em #1})}
\def\article{\@ifnextchar[{\earticle}{\oarticle}}
\def\oarticle#1#2#3#4#5#6{{\rm #1}, {\ital ``#6''}, {\rm #2 #3 (#5) #4}}
\def\earticle[#1]#2#3#4#5#6#7{{\rm #2}, {\ital ``#7''}, {\rm #3 #4 (#6) #5}  [\hhrefq{#1}]}
\def\hepart[#1]#2{{\rm #2, \sl#1}}
\def\heparticle[#1]#2#3{#2, {\ital ``#3''} [\hhrefq{#1}]}
\newcommand{\doi}[1]{\href{http://dx.doi.org/#1}{[link]}}

\newcommand{\hhrefqq}[1]{\IfBeginWith{#1}{10.}{\href{https://doi.org/#1}{doi:#1}}{\hhrefq{#1}}}
\def\earticle[#1]#2#3#4#5#6#7{{\rm #2}, {\ital ``#7''}, {\rm #3 #4 (#6) #5}  [\hhrefqq{#1}]}

\renewenvironment{thebibliography}[1]
{\begin{multicols}{2}[\section*{\refname}]%
		\@mkboth{\MakeUppercase\refname}{\MakeUppercase\refname}%
		\list{\@biblabel{\@arabic\c@enumiv}}%
		{\settowidth\labelwidth{\@biblabel{#1}}%
			\leftmargin\labelwidth
			\advance\leftmargin\labelsep
			\@openbib@code
			\usecounter{enumiv}%
			\let\p@enumiv\@empty
			\renewcommand\theenumiv{\@arabic\c@enumiv}}%
		\sloppy
		\clubpenalty4000
		\@clubpenalty \clubpenalty
		\widowpenalty4000%
		\sfcode`\.\@m}
	{\def\@noitemerr
		{\@latex@warning{Empty `thebibliography' environment}}%
		\endlist\end{multicols}}

\newcounter{alphaequation}[equation]
\def\thealphaequation{\theequation\hbox to
	0.6em{\hfil\alph{alphaequation}\hfil}}
\def\eqnsystem#1{
	\def\@eqnnum{{\rm (\thealphaequation)}}
	\def\@@eqncr{\let\@tempa\relax \ifcase\@eqcnt \def\@tempa{& & &} \or
		\def\@tempa{& &}\or \def\@tempa{&}\fi\@tempa
		\if@eqnsw\@eqnnum\refstepcounter{alphaequation}\fi
		\global\@eqnswtrue\global\@eqcnt=0\cr}
	\refstepcounter{equation} \let\@currentlabel\theequation \def\@tempb{#1}
	\ifx\@tempb\empty\else\label{#1}\fi
	\refstepcounter{alphaequation}
	\let\@currentlabel\thealphaequation
	\global\@eqnswtrue\global\@eqcnt=0 \tabskip\@centering\let\\=\@eqncr
	$$\halign to \displaywidth\bgroup \@eqnsel\hskip\@centering
	$\displaystyle\tabskip\z@{##}$&\global\@eqcnt\@ne
	\hskip2\arraycolsep\hfil${##}$\hfil& \global\@eqcnt\tw@\hskip2\arraycolsep
	$\displaystyle\tabskip\z@{##}$\hfil
	\tabskip\@centering&\llap{##}\tabskip\z@\cr}
\def\endeqnsystem{\@@eqncr\egroup$$\global\@ignoretrue} \makeatother

\definecolor{Gray}{gray}{0.95}

\def\bal#1\eal{\begin{align}#1\end{align}}

\usepackage{tocloft}
\begin{document}
\thispagestyle{empty}
\begin{center}  
{\LARGE\bf\color{rossos} Dark Matter as dark dwarfs and\\[0.3em]  other macroscopic objects: multiverse relics?} \\
\vspace{0.6cm}
{\bf Christian Gross$^{a}$, Giacomo Landini$^{a}$, Alessandro Strumia$^{a}$, Daniele Teresi$^b$}\\[6mm]

{\it $^a$ Dipartimento di Fisica, Universit\`a di Pisa, Pisa, Italia}\\[1mm]
{\it $^b$ CERN, Theoretical Physics Department, Geneva, Switzerland}\\[1mm]

\vspace{0.5cm}
{\large\bf Abstract}
\begin{quote}\large
First order phase transitions can leave relic 
pockets of false vacua and their particles, that manifest as macroscopic Dark Matter. 
We compute one predictive model: 
a gauge theory with a dark quark relic heavier than the confinement scale.
During the first order phase transition to confinement, dark quarks remain in the false vacuum and
get compressed, forming Fermi balls that can undergo gravitational collapse to stable dark dwarfs
(bound states analogous to white dwarfs) near the Chandrasekhar limit, or primordial black holes.

\end{quote}
\end{center}
\setcounter{page}{1}
{\small\tableofcontents}
	
\section{Introduction}
Dark Matter (DM) might be an accidentally stable dark baryon made of dark quarks $q$ colored under a new dark gauge group~\cite{BaryonicDM}.
In models with an appropriate number of light dark quark flavours 
the dark confinement phase transition is first-order and has interesting cosmological implications~\cite{Witten:1984rs,1810.04360}:
relic dark quarks tend to remain in the false vacuum (because {they are} lighter than dark baryons in the true vacuum),
so expanding bubbles of the true vacuum compress them down to small pockets.
In the presence of a dark asymmetry this process {can lead} to macroscopic DM relics~\cite{1810.04360}
(which could also be in a color superconducting phase as we point out).

\smallskip

A similar first-order phase transition takes place in models with no light dark quarks.
Heavy relic dark quarks remain in the false vacuum because they cannot access the confined phase as free quarks
(until they meet and form dark baryons) and get compressed to small pockets.  

If dark quarks are only mildly heavy,
such pockets evaporate leaving no macroscopic remnants when dark baryon formation occurs~\cite{2103.09827}.

\smallskip

If relic dark quarks are heavy enough that their gravity becomes relevant, 
after the initial stage of compression, a gravitational collapse can take over and lead to a new kind of macroscopic DM relic.
This is one of the main new points of this paper.

Depending on the dark quark mass $m$, pockets can form stable relic {\em dark dwarfs}
(acceptable DM candidates analogous to white dwarfs, stabilized by quantum pressure against gravity) or {\em black holes} (that evaporate if light enough,
or remain as possibly acceptable DM candidates if heavy enough).\footnote{Our mechanism is different from black hole production via collisions of bubbles (see e.g.~\cite{bubblecollisision}) and from the fermion soliton stars and black holes discussed in \cite{LeePang}.}
Dark scalar quarks only form black holes.
In section~\ref{1storder} we summarize when a strongly interacting gauge theory gives a first-order confinement phase transition.
In section~\ref{pretransition} we discuss the phase transition in our model with no light quarks.
In section~\ref{pockets} we discuss the subsequent gravitational collapse of surviving pockets in the unconfined phase.

\smallskip

In the final section~\ref{DMmulti} we discuss the possibility that the above phenomenon, 
studied in the context of first-order phase transitions in strongly interacting gauge theories, is more general.
In the multiverse context, scalars might give post-inflationary first-order phase transitions among false vacua down to the SM vacuum.
Particles which are lighter in a false vacuum than in the SM vacuum could get trapped so that 
pockets of false vacua and their compressed light particles could survive within our universe, and be its  Dark Matter.
In such case, finding Dark Matter in possibly macroscopic pockets of false vacua
would allow to explore the multiverse  beyond our vacuum.

\smallskip

Conclusions are given in section~\ref{concl}.

\section{First-order phase transitions from strong interactions}\label{1storder}
To start, we here summarize how first-order phase transitions to confinement
arise in strongly-interacting gauge theories with (section~\ref{LQ}) and without (section~\ref{HQ}) light quarks.

\subsection{Strongly-interacting gauge theories with light quarks}\label{LQ}
A key element of the scenario is a first-order phase transition. 
We consider
a non-Abelian gauge group $G$ with  $N_f$ flavours of Dirac fermionic quarks lighter than the confinement scale $\Lambda$.
We focus on $G=\SU(N)$, so that it is known when non-perturbative gauge interactions give a first order confinement phase transition~\cite{Svetitsky:1982gs,hep-lat/0307017,hep-lat/0502003,1204.6184}:
\beq N_f=0 \qquad\hbox{or}\qquad 3\le N_f \circa{<}3N .\eeq
For $N_f >0$ the order of the phase transition can be  computed analytically from coefficients of RG equations in the pion effective theory~\cite{Pisarski:1983ms},
as well as from lattice simulations.
$N_f=0$ is special because it leads to no pions, and only lattice simulations are available.

The possibility with $N_f>0$ light quarks has been studied in~\cite{1810.04360}, that we briefly summarize.
At $T\sim \Lambda$ bubbles of the true vacuum appear and expand.
Quarks in the false deconfined vacuum are lighter than hadrons in the confined true vacuum.
So quarks can only partially cross the bubble walls, and tend to be compressed in the surviving pockets of false vacuum.
Assuming a dark baryon asymmetry $Y$, such pockets contain $Q \sim Y (R_i \Lambda)^3 \gg 1$ dark quarks.
Here $R_i$ is the initial radius of pockets, estimated to be of order 
$R_i \sim M_{\rm Pl}^{2/3}/\Lambda^{5/3} $,
where $M_{\rm Pl}\approx 1.2~10^{19}\GeV$ is the Planck mass.
Compression leads to balls of dense matter, stabilised by Fermi pressure, with radius $R\sim Q^{1/3}/\Lambda$.
For appropriate values of $N$ and $N_f$, such pockets are stable because they are lighter than free hadrons.
Macroscopic objects with super-Planckian mass are easily obtained:
approximate predictions in the ($M,R$) plane are plotted in fig.\fig{MacroCompositeDM}, where we also consider the possibility of bosonic quarks,
stabilised by their quantum pressure.
We will mention a new possibility in section~\ref{pockets}: pockets of light dark quarks in the color superconducting phase.

\medskip

Self-gravity of pockets is negligible here, as light dark quarks have the same density of dark gluons.
It will be important in the other case: $N_f=0$ light dark quarks and a heavy one.

\subsection{Strongly-interacting gauge models with heavy quarks}\label{HQ}
The other possibility, no light quark flavour, has not been discussed in ~\cite{1810.04360}.
Heavy free dark quarks, being heavier than Coulombian dark baryons made of them, would not give rise to 
pockets stabilised by dark strong interactions.
In this paper we will show that gravity can stabilise pockets.
In order to have an asymmetry, we assume the presence of one heavy flavour of dark quarks $q$ with mass $m\gg \Lambda$.
The theory is
\beq\label{eq:LDM}
\Lag = \Lag_{\rm SM}- \frac14 {G}_{\mu\nu}^a G^{\mu\nu a} + \bar q (i\slashed{D}-m) q \eeq
where the omitted dark topological term plays no role, and the dark gauge coupling runs as
\begin{equation}\label{eq:RGEsol}
\alpha_{\rm dark}(E) \approx \frac{6\pi}{11N} \frac{1}{\ln E/\Lambda}.
\end{equation}
We denote as $T_{\rm dark}$ the temperature of the dark sector, and as $T_{\rm SM}$ the possibly different temperature of
the SM sector.
We allow for the possibility that the two sectors are
negligibly coupled, and can thereby have different temperatures.
We define  \beq r = \rho_{\rm dark}/\rho\eeq 
as the fraction of total energy $\rho=\rho_{\rm SM}+\rho_{\rm dark}$
in the dark sector, evaluated at the critical temperature $T_{\rm cr}$ of the dark confinement phase transition.

\smallskip

We have two possible DM candidates: dark baryons and dark glue-balls.

We mostly consider dark heavy quarks with number density $n = Y s_{\rm dark}$ assumed to be
dominated by a (possibly small) asymmetry $Y$.
Here $s_{\rm dark}=2\pi^2 g_{\rm dark} T^3_{\rm cr}/45$ is the entropy of the dark sector,
and $g_{\rm dark}=2 (N^2 - 1)$ is its number of dark gluon degrees of freedom.

\smallskip

We are here interested in a new generic phenomenon that can happen with dark baryons,
so we only say a few words about dark glue-balls.
They tend to be long-lived in models with no light dark quarks.
If stable enough, dark glue-balls can be acceptable DM candidates provided that 
the dark sector temperature is initially much smaller than the
SM temperature,  for example because the dark sector is populated via gravitational freeze-in~\cite{2012.12087}.
In such a case, after confinement dark glue-balls undergo
`cannibalistic' $3\leftrightarrow 2$ processes~\cite{cannibalism,1602.04219} that decouple when 
$T_{\rm dark}= T_{\rm dec} \approx T_{\rm cr}/3\ln x \approx T_{\rm cr}/25$ 
where $x\approx \Ord(1) (M_{\rm Pl}/T_{\rm cr})^{1/4}g_{\rm dark}^{1/24}r^{1/8}$. 
Taking into account that comoving entropy is separately conserved in each sector (SM, dark gluons, dark quarks), the
cosmological DM abundance is reproduced if\footnote{
The final DM abundance is different in models where dark glue-balls instead decay, possibly injecting significant entropy;
 for example the dark gauge group might be unified with a part of the SM gauge group at a scale not much above $\Lambda$.}
\beq  \label{eq:YDM}
f Y m  + T_{\rm dec} \approx 0.4\eV   \frac{g_{\rm SM} T_{\rm SM}^3} {g_{\rm dark} T_{\rm dark}^3} \eeq
where the second contribution comes from dark gluons, and the first contribution from dark quarks.
In our context, their initial abundance can get reduced by a factor $f\le 1$ if black-holes form and evaporate
converting dark quarks inside into radiation.
Such effects arise because the first-order dark confinement phase transition leads to qualitatively new features, as we now discuss.

\medskip

Dark quarks, despite being massive, 
can  enter the confined region only if they find other dark quarks to form dark baryons~\cite{Witten:1984rs,2103.09827}.
Compression of relics much heavier than the rest of the Universe results in a higher density,
and ultimately into a gravitational self-attraction that can form compact objects, leaving gravitational relics. Indeed, the potential energy of a pocket with radius $R$ at low temperature is 
\beq \label{eq:Unr} 
U(R) =\Delta V \, \frac{4\pi R^3}{3} + \sigma \, 4\pi  R^2 +
\frac{9}{20}\left(\frac{3\pi^2}{2 N^2}\right)^{1/3}
\frac{Q^{p}}{mR^2} -\frac{3Q^2 m^2}{5 R M_{\rm Pl}^2} .\eeq
The first term is vacuum energy; the second term is the wall energy;
the third term  is the quantum pressure (with $p=5/3$ for a non-relativistic fermion, and $p=1$ for a non-relativistic boson);
the last term accounts for gravity.\footnote{Order one coefficients 
are here computed for a fermionic dark quark sphere with uniform density and weak gravity.
We will later consider the realistic case with deviations from this.
For a boson with $g_{\rm boson}$ degrees of freedom
the coefficient in front of its quantum pressure becomes $g_{\rm boson} \pi^2$.
In the boson case an extra term that accounts for short-range dark nuclear interactions can be relevant.}
As well known, it allows for non-trivial minima of $U(R)$.

\section{The pre-transition bubbles phase}\label{pretransition}
The phase transition to confinement that happens in the dark sector is of first order.
At the critical temperature $T_{\rm dark} = T_{\rm cr} \sim\Lambda$
the two phases coexist, as a bubble of one vacuum within the other vacuum
is kept in equilibrium by energetic plus entropic forces. 
This is formally described by degeneracy of the effective thermal `potential', $\Delta V_T=0$.
At lower temperature, large enough bubbles of the true vacuum expand in the false vacuum releasing a latent heat density ${\cal L}$.
In our situation ${\cal L}$ is positive because the low-temperature phase is more ordered than the high-temperature phase.
Furthermore ${\cal L}$ is significant because thermal effects grow with couplings, and we are at strong coupling.

The universe super-cools below the critical temperature, and expanding bubbles of the confined phase start nucleating.
In the thin-wall approximation bubbles have surface tension $\sigma$ and
appear with initial radius $R_{\rm cr} = 2 \sigma/\mathcal{L} \delta$. The
space-time density of nucleations is~\cite{2103.09827}
\beq \gamma\approx T_{\rm cr}^4 \exp \left[-\frac{\kappa }{\delta^2}\right],\qquad
\hbox{where}\qquad
\delta = 1- \frac{T_{\rm dark}}{T_{\rm cr}},\qquad
\kappa =\frac{16\pi}{3}\frac{\sigma^3}{{\cal L}^2 T_{\rm cr}}.
\eeq
According to lattice computations performed for the SU(3) gauge group~\cite{hep-lat/0502003}, 
the critical temperature is $T_{\rm cr} \approx\Lambda$,
the latent heat density is ${\cal L} \approx 1.4\, T_{\rm cr}^4 $, 
and the wall surface tension is $\sigma \approx 0.02 T_{\rm cr}^3$.
So $\kappa \approx 0.7~10^{-4}$ is small, and the 
exponential suppression is lost when $\delta^2 \sim \kappa$, after little super-cooling~\cite{Witten:1984rs} (which justifies the thin-wall approximation).
According to lattice simulations, $\kappa$ is similarly small at least up to $N \circa{<}10$, as
${\cal L} \approx (0.76 - 0.3/N^2)^4 N^2 T_{\rm cr}^4$,
$\sigma \approx (0.015 N^2-0.1)T_{\rm cr}^3$
($\sigma$ might instead grow linearly with $N$)~\cite{hep-lat/0502003}.

\subsection{Calculation of the distance between bubbles, $R_0$}\label{R0}
The average distance between bubbles, $R_0$, can be computed as follows.
Since latent heat is significant, nucleation and expansion of bubbles reheats the dark sector: 
this slows bubble walls and blocks nucleation of new bubbles.
This happens when the fraction $x$ of the Universe volume
in the confined phase is large enough that its released latent heat
reheats the rest of the universe up to almost the critical temperature $T_{\rm cr}$:
\be\label{eq:delta01}
x_0 \approx\frac{4 \pi^2 g_{\rm dark}\, T_{\rm cr}^4 }{30 {\cal L}}  \, \delta_0 \,.
\ee
We ignored dark quarks, assuming that, at this initial stage, their energy density is much smaller than the
energy density of dark gluons.
When the phase transition starts, the growing $x\ll 1$ can be approximated, up to order one factors, as
\be\label{eq:delta0relation}
x_0 \approx \int _0^{t(\delta_0)}\frac{4 \pi [G(\delta_0,\delta) R_{\rm cr}]^3}{3} \gamma(\delta) \, dt \approx \frac{4 \pi [G(\delta_0,\delta_p) R_{\rm cr}]^3}{3}    \gamma(\delta_p) \frac{ \delta_p }{ H_{\rm cr}}
\ee
where $H_{\rm cr} = \sqrt{\sfrac{4\pi^3 g_{\rm dark}}{45 r}}\, \sfrac{T_{\rm cr}^2}{M_{\rm Pl}}$ is 
the Hubble rate at the critical temperature.
Indeed the nucleation rate $\gamma$ is exponentially sensitive to $\delta$.  
If this were the dominant factor, the integral would be dominated by $\delta_0$.
However, it is dominated by a mildly earlier $\delta_p$, because the growth factor $G$ of bubble radii, 
despite being only polynomial, is enhanced by an $M_{\rm Pl}/T_{\rm cr}$ factor
\be
G(\delta_0,\delta_p) \approx  1 + \frac{v(\delta_0)(t-t_p)}{R_{\rm cr}(\delta_p)} \approx  
\frac{\epsilon \delta_0 (\delta_0-\delta_p)}{H_{\rm cr}  R_{\rm cr}(\delta_p)} .
\ee
As discussed in the next section,
bubble walls move with speed $v \approx \epsilon \delta$, possibly suppressed by a mild 
Boltzmann factor  $\epsilon \sim e^{-M_{\rm DG}/T_{\rm cr}} \approx e^{-6}$, 
in models where heat exchange between the two phases only proceeds through dark glueballs. 
Equating eq.\eq{delta0relation} with eq.\eq{delta01} gives
\be
 \delta_p \approx \sqrt{
\frac{\kappa}{\ell}}\approx 0.001,
\qquad x_0 \approx 0.1 \, \frac{g_{\rm dark}}{100} ,\qquad \ell = \ln \[ \frac{10125 r^2}{8 \pi^7 g_{\rm dark}^3} \frac{\mathcal{L}}{T_{\rm cr}^4} \frac{M_{\rm Pl}^4}{T_{\rm cr}^4} \epsilon^3 \delta_p \delta_0^2 (\delta_0 - \delta_p)^3\]
\ee
with little dependence on parameters such as $T_{\rm cr}$ or $g_{\rm dark}$ that appear in the log; 
moreover, we can approximate $\delta_0 \approx \delta_p \approx \sqrt{\kappa}$ in the log. 

\smallskip

\begin{figure}[t]
$$\includegraphics[width=\textwidth]{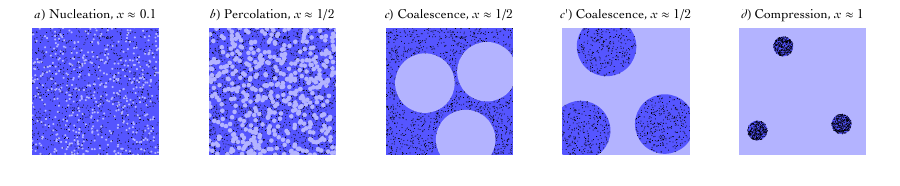}$$
\caption{\label{fig:Bolle}\em Sketch of the phase transition.
Blue: false vacuum; white: true vacuum; black dots: heavy quarks.
a) Bubbles appear.
b) Bubbles start merging;
c) Bubbles merged.
c$'$) The computation switches to the pockets approximation.
d) Pockets contract.
}
\end{figure}

The average distance $R_0$ between bubbles is thereby related to their
size $R(\delta_0) $ when nucleation stops as
\be\label{eq:R0expansion}
R_0 \approx x_0^{-1/3} R(\delta_0),\qquad
R(\delta_0) \approx v \Delta t \approx  \frac{\epsilon \delta_0^2}{H_{\rm cr}} .
\ee
In view of $\delta_0 \ll 1$ a Hubble volume contains many bubbles, as illustrated in fig.\fig{Bolle}a.

\subsection{Calculation of the distance between pockets, $R_1$}
The growth speed of bubbles is limited by the fact that the  latent heat released by bubble expansion
raises the
temperature of the false vacuum, and that bubbles only expand if $T_{\rm dark}< T_{\rm cr}$.
Their speed satisfies the bound $\dot R \circa{<}\delta$~\cite{Witten:1984rs,2103.09827}.\footnote{We assumed a homogeneous temperature,
neglecting possible warming around bubbles.
Furthermore, this general upper bound on bubble speed is stronger in our case: 
as we have no light quarks and thereby no light pions, 
bubbles must convert outside gluons into inside glue-balls to expand.
So this rate is suppressed by a mild Boltzmann factor that makes walls slower,
$v\sim e^{-M_{\rm DG}/T_{\rm cr}}$.
Indeed the glue-ball mass is $M_{\rm DG}\sim 6 \Lambda$  
and
 gluons at $T_{\rm cr}$ are presumably lighter than glue-balls.
At leading order, the thermal mass of a vector is $m_V^2 = g^2 T^2 N/6$ 
in the absence of matter lighter than $T$ (see e.g.~\cite{1310.6982}).
This mild Boltzmann suppression is avoided assuming that the DM and SM sectors interact.
We anyhow assume that the speed limit on bubble velocity is sub-dominant with respect to
the bound discussed next.
}
The system then rapidly approaches an attractor solution, where $T_{\rm dark}$ 
stays at the special value just below $T_{\rm cr}$ such that released latent heat is compensated by Hubble expansion
\be  0\approx \frac{d T_{\rm dark}}{dt} = - H T_{\rm dark} +  \frac{{\cal L} ~ dx/dt}{d \rho_{\rm dark} / d T_{\rm dark} } 
\approx - H T_{\rm dark} +    \frac{1.1 T_{\rm cr}}{g_{ \rm dark}} \frac{d x}{dt}.
\ee
So the volume fraction in the true vacuum, $x$, grows linearly with time:
$\dot x \approx  g_{\rm dark} H $, having inserted order one numerical factors appropriate for $\SU(3)$.
The time needed to fill about half of the space reaching $x=x_{\rm perc} \approx 1/2$ is a fraction of a Hubble time
independently of $T_{\rm SM}/T_{\rm dark}$:
\beq t_{\rm perc} \approx  \frac{0.5}{g_{\rm dark} H}.\eeq
We neglected $x_0$ compared to $x_{\rm perc}$
and the fact that $g_{\rm dark}$ changes when
glue-balls become relevant.
At this `percolation time' bubbles start meeting while having average radius $R_0$.
The correction due to the overall Hubble expansion of the universe is small:
despite their slow non-relativistic velocity, bubbles merge faster than the Hubble rate because
there are many bubbles per Hubble volume, $N_{\rm bubble}\sim 1/(H R_0)^3 $.
This situation is plotted in fig.\fig{Bolle}b.

\smallskip

When bubbles collide a new phenomenon starts: 
coalescence of small bubbles into bigger ones.
The time needed for changing shape by moving the mass 
such that two bubbles with radius $R$ merge into one  bubble with bigger radius $2^{1/3}R$
is estimated as~\cite{Witten:1984rs,2103.09827}: 
\be
t_{\rm coal}(R) \approx \sqrt{ \frac{2 \pi^2 g_{\rm dark} T_{\rm cr}^3}{90 (2 - 2^{2/3}) \sigma}} T_{\rm cr}^{1/2} R^{3/2}.
\ee
At the beginning coalescence is fast and bubbles that touch with $R \approx R_0$ immediately form bigger bubbles.
Merging progresses and the size $R$ of bubbles grows exponentially.
At some point, when $R\approx R_1$, coalescence becomes slower than bubble growth, $t_{\rm coal}(R_1)\approx  t_{\rm perc}$.
We thereby obtain the radius of bubbles 
\beq R_1\approx 0.5  \sqrt[3]{\frac{2025(2-2^{2/3})r}{4\pi^5g_{\rm dark}}}\frac{\sigma^{1/3}}{g_{\rm dark}^{}}\frac{M_{\rm Pl}^{2/3}}{T_{\rm cr}^{8/3}}\approx \frac{0.12r^{1/3}}{g_{\rm dark}^{4/3}}\frac{M_{\rm Pl}^{2/3}}{T_{\rm cr}^{5/3}}.\eeq
This situation is plotted in fig.\fig{Bolle}c.

\section{The post-transition pockets phase}\label{pockets}

After bubbles have merged, the typical  size of the remaining big regions in the false vacuum is
\beq R_i \approx \max(R_0, R_1),\eeq
smaller than the horizon size $1/H_{\rm cr}$, which is also the Schwarzschild radius of the homogeneous universe.
The universe can now be approximated as being in the confined phase (true vacuum), up to remaining relic bubbles in the free phase (false vacuum).
These bubbles can be approximated as  spherical and dubbed `pockets' in order to avoid confusion with the bubbles of the condensed phase studied in the previous section.\footnote{Similar objects containing light quarks have been dubbed  `nuggets' in~\cite{Witten:1984rs}.
A possibly more appropriate name is `Asterix villages'  resisting to the compression by expanding Romans.
Then, our pockets containing heavy quarks could be dubbed `Obelix villages'.} 
This situation is plotted in fig.\fig{Bolle}c$'$, equivalent to fig.\fig{Bolle}c.

The pockets with initial radius $ R_i$ 
shrink compressing the relic dark quarks that cannot enter the confined region,
as long as particle-physics processes are negligible
(in section~\ref{DBF} we will show that dark-baryon formation is negligible). 
The number density of relic dark quarks at percolation, $x_{\rm perc}\approx 1/2$, is $n \approx Y s_{\rm dark}/x_{\rm perc}$,  and
the initial number of dark quarks in a pocket is $Q \approx n\,4\pi R_i^3/3\sim  Y (R_i  \Lambda)^3$.
The total excess mass of a pocket compared to the cosmological average is $M = Q m$.

\smallskip

The true-vacuum expansion described in section~\ref{R0} keeps going on, rephrased from the
old language (expansion of bubbles) to the new language (compression of pockets).
The compression speed remains limited by the rate at which the expanding Universe absorbs
the latent heat 
released during the compression, $W ={\cal L} \,\dot V$.
The steady state with $T_{\rm dark}$ very close to $T_{\rm cr}$ prevents formation of bubbles inside pockets
and proceeds until, after another time $\sim t_{\rm perc}$,
the isothermal compression reaches $R \ll R_i$ and most of the Universe is the true vacuum.
At this point the kinetic energy of walls is small enough that pockets do not get crunched and 
various new contributions to the pressure on the small pockets start becoming relevant.
We list such pressures according to how they scale with $R$, starting from those more important at larger $R$:
\begin{itemize}
\item[\color{red}$\triangleright\triangleleft$] The inward pressure due to {\bf latent heat} or (at $T \ll T_{\rm cr}$) {\bf vacuum energy},  
\beq p_V ={\cal L},\Delta V \propto R^0.\eeq
\item[\color{red}$\triangleright\triangleleft$] The inward pressure due to the {\bf wall} tension, $p_\sigma =2\sigma/R$.
This is negligible compared to $p_V$ for $R \circa{>} 1/\Lambda$.
\item[\color{blue}$\triangleleft\triangleright$] 
The outward pressure due to the {\bf thermal gas} of trapped dark quarks, $p_{\rm gas}=nT = Q T/V $ in the non-relativistic limit. This scales as $p_{\rm gas} \propto 1/R^3$ if $T$ is constant (see later).\footnote{The outward pressure due to gluons is included in $p_V$, with $V$ being the  potential at finite temperature and zero chemical potential. The full dynamics could also be studied using finite temperature and density, 
considering the Landau potential.
We prefer to separately include the pressure due to quarks.}

\begin{figure}[t]
$$\includegraphics[width=\textwidth]{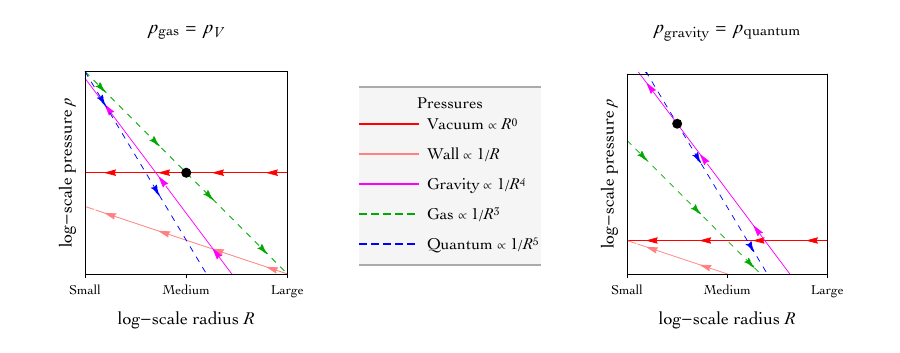}$$
\caption{\label{fig:PressioniPocket}\em Radius dependence of the various
contribution to total pressure on pockets.
Dashed pressures with left-wing arrows tend to expand pockets.
Continuous pressures with right-wing arrows tend to compress pockets.
Left: usual case where gas pressure is more relevant than the pressure due to gravity.
Right: gravity more relevant than gas.
}
\end{figure}

\item[\color{red}$\triangleright\triangleleft$] The inward pressure due to {\bf gravitational} attraction inside the pocket.
The gravitational energy is $U_{\rm grav} \sim -G_{\rm N} M^2/R$ in the Newtonian limit, 
and thereby \beq p_{\rm grav} \sim  -U_{\rm grav}/R^3 \sim  Q^2 m^2/R^4 M_{\rm Pl}^2.\eeq

\item[\color{blue}$\triangleleft\triangleright$] The outward {\bf quantum} pressure, approximated by
\beq   \begin{array}{c|c|c}
p_{\rm quantum} & \hbox{Dirac fermion} & \hbox{Boson}  \\ \hline
\hbox{non-relativistic}  &\displaystyle   {\frac{9}{40}\(\frac{3}{2\pi N^2}\)^{1/3}} \frac{Q^{5/3}}{mR^5} \sim \frac{n^{5/3}}{m} &\displaystyle  {\frac{g_{\rm boson} \pi}{2}} \frac{Q}{mR^5}\\[2mm]
\hline
\hbox{relativistic} & \displaystyle  {\frac{3}{16}\(\frac{9}{4\pi^2 N}\)^{1/3}} \frac{Q^{4/3}}{R^4} \sim n^{4/3}&\displaystyle {\frac{g_{\rm boson}}{4}}  \frac{Q}{R^4}\\
\end{array}
\eeq
where order one factors assume constant density.
In the fermionic case the Fermi pressure 
$p_{\rm Fermi} \sim n K$ arises 
because fermions with number density $n \sim Q/R^3 $ fill energy levels up to the
Fermi momentum $k \sim   n^{1/3}$, that corresponds to kinetic energy $K=k^2/2m$
(non-relativistic) or $K = k$ (relativistic).
The Fermi pressure can be written in terms of $n$, and is thereby an intensive quantity.

In the bosonic case all quanta with $g_{\rm boson}$ degrees of freedom can fill the lowest-energy states, with momentum $k \sim 1/R$ 
for an object of size $R$. The bosonic pressure is thereby a smaller finite-size effect, similar to Casimir energy,
that can be sub-dominant with respect to effects due to interactions.

\item[\color{blue}$\triangleleft\triangleright$] {\bf Interactions} among quarks could give larger effects than the bosonic pressure.
Enhanced long-range interactions arise if dark quarks are charged under some Abelian gauge interaction (such as electromagnetism):
a pocket containing a quark asymmetry is subject to a Coulomb pressure
$p_{\rm Coulomb} \sim \alpha Q^2/R^4$, outward because like charges repel.
Our non-Abelian dark gauge interactions generate no such pressure, as two dark quarks can attract or repel.
Thereby formation of dark baryons is not enhanced by $Q^2$.
Short-range particle physics processes can lead to formation of dark baryons or annihilation of dark quarks.
For the moment we neglect such possible effects, to be discussed in section~\ref{DBF}.

\end{itemize}
The final stage of the compression proceeds with constant small speed~\cite{2103.09827}.
One might worry that compression heats the pocket, triggering reactions inside.
We now argue that the pocket temperature tends to remain close to $T_{\rm cr}$.
The energy flow is approximated by the Stefan-Boltzmann law times a suppression $\epsilon$
\beq \label{eq:Enflow}W_{\rm rad} = \frac{4\pi   R^2}{120} ~ \pi^2 (T_{\rm in}^4 \epsilon_{\rm in} - T_{\rm out}^4\epsilon_{\rm out}) .\eeq 
As discussed in the previous section, we expect $\epsilon_{\rm in} \sim e^{-M_{\rm DG}/T_{\rm in}} $, as
gluons inside with energy $\sim T_{\rm in}$ must become glue-balls outside with mass $M_{\rm DG}$.
The same factor $\epsilon_{\rm out} \sim e^{-M_{\rm DG}/T_{\rm out}}$
arises for the flux going into the pocket, as the dark glue-ball density outside is Boltzmann suppressed.
In view of this large exponential factor, the temperature inside tends to stay roughly constant at $T \sim T_{\rm cr}$:
 temperatures higher than $T_{\rm cr}$ cool easily. Cooling of pockets below $T_{\rm cr}$ needs an exponentially slow time
if the dark sector negligibly interacts within the SM sector.

 \subsection{Possible final states, ignoring gravity}
Fig.\fig{PressioniPocket}a shows how the pressures depend on radius.
The pockets can evolve in different ways depending on which contributions to the pressure on the walls dominates.
For the moment we assume that gravity is negligible, and summarize what can happen:
\begin{itemize}
\item[a)] {\bf Thermal balls}. 
In the case plotted in fig.\fig{PressioniPocket}a, 
the thermal pressure of dark quarks trapped inside pockets can temporarily stop their compression, $p_{\rm gas} = p_{V}$, while all other pressures are negligible.
Then, pockets reach a minimal radius $R^{\rm min}_{\rm gas} \approx 2 R_i (Y T/\Lambda)^{1/3}$
where $T \sim T_{\rm cr} \approx \Lambda$.

Later, quarks can leak out, the Universe cools down, the pockets slowly cool and compress further.
As pockets get compressed more, different things can happen.

 \item[b)] {\bf Nothing}.
 One possibility is that pockets evaporate because dark quarks either annihilate with anti-quarks, or
(in models with quarks only) form dark baryons, that leak out~\cite{2103.09827}.
Furthermore, cold pockets can be destroyed by bubbles that form inside.

\item[c)] {\bf Fermi or Bose balls}.
Alternatively, particles in the pocket might have no way to escape,
and compression can slowly proceed up to when 
pockets get stabilised by quantum pressure, while gravity remains negligible.
In some models with light dark quarks, $m \circa{<}\Lambda$, 
strong dynamics makes it energetically favourable for baryons to stay inside pockets~\cite{1810.04360}.
A similar situation can happen in models with ad-hoc first order phase transitions~\cite{1301.0354,2008.04430}.
In general, relics remain if friction keeps walls non-relativistic and if trapped particles are enough heavier outside than inside, so
that the relativistic quantum pressure inside gives $p_{\rm quantum}=p_{V}$
with radius $R\sim Q^{p/4}/\Lambda$, where $p=1$ for bosons, $p=4/3$ for fermions.
 
 \item[d)]
 A new possibility that can happen in strongly-interacting models with light quarks is that a new phase, known as
{\bf Color Superconductivity}~\cite{hep-ph/9711395}, exists at large density.
In this phase $\langle qq \rangle$ condensates break the dark color gauge group
and the approximate accidental global symmetries. Therefore, the equation of state of dark quark matter would be consequently modified  compared to the c). For three light dark flavours, a color-flavor locking phase~\cite{0709.4635}, which leaves an unbroken global $\SU(3)$ symmetry, would be favored.\footnote{We thank Michele Redi for pointing out this possibility.}

 \end{itemize}

 \subsection{Possible final states, including gravity}\label{fingrav}
A new possibility arises in models with sufficiently heavy quarks:
the pockets with initial radius $R_i$
may get compressed so much that, at some point,
the inward pressure $p_{\rm grav}$ due to gravity becomes larger than the pressures $p_{\rm gas}, p_{V}$ 
assumed to be dominant so far, see fig.\fig{PressioniPocket}a.
If instead the number $Q $ of heavy quarks with mass $m$ inside pockets is large enough that
\beq \label{eq:Qcr}p_{\rm grav} \circa{>} p_{\rm gas}=p_V\qquad \hbox{i.e.}\qquad
Q \circa{>}  Q_{\rm cr}\approx  0.1 \frac{M_{\rm Pl}^3}{m^3} \frac{T^2}{T_{\rm cr}^2} \eeq
we are in the situation of fig.\fig{PressioniPocket}b, and a gravitational collapse happens.
For later convenience, we notice that for $T \sim T_{\rm cr}$ this is parametrically the same as the Chandrasekhar condition.
Assuming that no heavy dark quarks exit from the pocket, their number is given by $Q\sim n \,4\pi R_i^3/3$ with  $n \approx 2 Ys_{\rm dark}$.

In order to keep formul\ae{} simple, from now on we assume that heavy quarks alone reproduce the total DM density, 
and that it is given by eq.\eq{YDM} in the limit $T_{\rm dec}=0$,
such that the contribution of dark glue-balls is negligible.
 With this assumption, the condition in eq.\eq{Qcr} for immediate gravitational collapse at $T=T_{\rm cr}$ is satisfied if dark quarks are heavier than 
\beq \label{eq:mcr}
m\circa{>} m_{\rm cr} \approx \min \left(  \frac{10^7 T_{\rm cr}^{3/2} }{\eV^{1/2} r^{3/8}},
 \frac{300  T_{\rm cr}}{r^{1/8}}\sqrt{ \frac{M_{\rm Pl}}{\eV}}\right)\eeq
which is plotted fig.\fig{DwarfMR}b and is sub-Planckian for low enough $T_{\rm cr}$.
For simplicity, we here set $N=3$ dark colors.
The corresponding homogeneous pre-compression energy densities of dark quarks and dark gluons
is $\sfrac{\rho_q }{\rho_g} \sim \sfrac{mY}{\Lambda} \sim \eV/m \ll 1$,
which justifies our assumption of neglecting heavy quarks in section~\ref{pretransition}.

\smallskip

Once the gravitational collapse starts, $p_{\rm gas}$ cannot prevent further compression, because
$p_{\rm grav}$ has a stronger dependence on $R$,
$p_{\rm grav} \propto -1/R^4$ while $p_{\rm gas} \propto 1/R^3$.
Dark quarks get closer than $1/\Lambda$ when the vacuum energy pressure $p_V \propto R^0$ is no longer relevant:
independently of the possible survival of the higher vacuum, matter can remain trapped by gravity.
For the moment, we keep considering the simpler case where 
the number of heavy quarks inside the pocket stays constant, while
the heat due to the collapse is radiated away as gluons and glue-balls (or, depending on the model, as SM particles), cf. eq. \ref{eq:Enflow}.
\smallskip

Two final states are possible: dark dwarfs or black holes.
\begin{itemize}
\item[e)] {\bf Dark dwarfs}. Gravitational collapse
proceeds up to when quantum pressure becomes relevant.
Non-relativistic Fermi pressure stops the gravitational collapse giving a dark analogous of 
white dwarfs, that we call ``dark dwarf''.
Minimising the quantum plus gravitational terms in eq.\eq{Unr} gives a radius 
that {decreases} with mass as
\beq R_{\rm dwarf} \approx \(\frac{81\pi^2}{16 N^2}\)^{1/3}
 \frac{M_{\rm Pl}^2}{m^3 Q^{2-p}} \sim\frac{M_{\rm Pl}^2}{m^{1+p} M^{2-p}}.\eeq
where $p=5/3$ for a fermion, $p=1$ for a boson.
The order unity numerical factor holds for a fermion in approximation of constant density and weak gravity.
The key new point is that when finally the
condensed phase fills the pocket, baryons remain trapped by gravity.

\begin{figure}[t]
$$\includegraphics[width=0.45\textwidth]{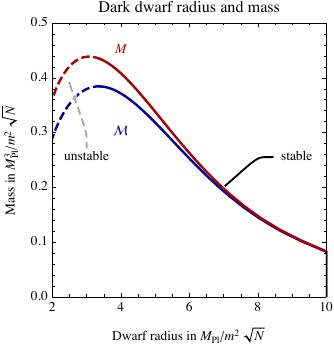}\qquad
\includegraphics[width=0.45\textwidth]{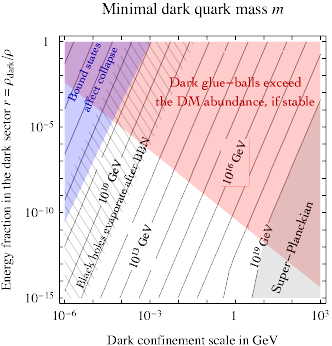}$$
\caption[Macroscopic DM]{\label{fig:DwarfMR}\em {\bf Left}: mass $M$ and gravitational mass ${\cal M}$ seen from outside 
as function of the radius of a dwarf formed with $N$ colors of a free fermion with mass $m$.
{\bf Right}: Minimal dark quark mass $m$ as function of the dark confinement scale 
and of $r = \rho_{\rm dark}/\rho$, for $N=3$ and $\epsilon=1$,
such that objects made of dark quarks collapse gravitationally and
reproduce the cosmological DM abundance in minimal cosmology.
Dark glueballs are extra DM candidates, if stable: in the shaded red region they exceed the DM abundance.
In the shaded blue region, bound-state formation during the gravitational collapse can modify it.
In the hatched region black holes 
evaporate between BBN and now, so their possible abundance is constrained.
}
\end{figure}

\item[f)] {\bf Dark black holes} form if $R_{\rm dwarf} \circa{<} R_{\rm Sch} = 2 M/M_{\rm Pl}^2$.
Let us discuss what this means:

\begin{itemize}
\item  In the bosonic case this condition implies relativistic momentum $k \sim1/R \sim m$,
so that the critical number of dark quarks that leads to black hole formation is $Q_{\rm BH} \sim (M_{\rm Pl}/m)^2$ (i.e.\ $M =Qm \sim M_{\rm Pl}^2/m$)
exceeded by eq.\eq{Qcr}. So bosonic quarks form black holes.

\item In the fermionic case this condition implies a relativistic Fermi momentum, $k_F \sim m$,
radius $R \sim M_{\rm Pl}/m^2$ and $Q \sim (M_{\rm Pl}/m)^3$ (i.e.\
$M = Qm \sim M_{\rm Pl}^3/m^2$).
So the condition for forming black holes is parametrically the same as the condition for forming dark dwarfs, eq.\eq{Qcr}.
\end{itemize}
\end{itemize}
Then, order one numbers are needed to understand if dark dwarfs or black holes form.
A precise computation is done using the 
Tolman-Oppenheimer-Volkoff (TOV) equations~\cite{Oppenheimer:1939ne,astro-ph/0605724}
for spherical hydrostatic equilibrium in general relativity
\beq 
\frac{dp}{dr} = - \frac{G}{r^2} \frac{({\cal M}+4\pi r^3 p)(\rho+p)}{1-2G{\cal M} /r},\qquad \frac{d{\cal M}}{dr}=4\pi r^2 \,\rho
\eeq
with boundary conditions $\rho(R)=0$ and ${\cal M}(0)=0$.
In our case, the equation of state is well approximated by $N$ dark colors of free fermions with mass $m$.
It can be parameterized in terms of the Fermi momentum $k(r)$ as
\beq \rho = \rho_0 (\sinh t -t),\qquad p = \frac{\rho_0}{3}(\sinh t  -8 \sinh\frac{t}{2} + 3t)\eeq
where
\beq \rho_0 = \frac{\pi N m^4}{4(2\pi \hbar)^3},\qquad
\qquad t = 4\ln \bigg(\frac{k}{m} + \sqrt{1+\frac{k^2}{m^2}}\bigg).\eeq
We here explicitly kept $\hbar=1$ to show that $N\neq 1$ can be compensated by a change of units
using the known TOV result (computed for neutron stars in the ideal limit of free neutrons, $N=1$).
We recomputed it, because unlike TOV we are not interested in the mass ${\cal M}$ as seen from outside gravity.
We are interested in the mass $M =Qm= \int_0^r dr\, 4\pi r^2 \,\rho/\sqrt{1-2G{\cal M}(r)/r}$.
Because of gravitational binding energy, ${\cal M}$ is smaller than $M$.
Our computation shown in fig.\fig{DwarfMR}a finds that
$M$ is $14\%$ higher than ${\cal M}$ at the threshold for black hole formation.
Thereby the bound on the number $Q = M/m$ of dark quarks is
\beq Q> Q_{\rm BH} = \frac{0.44}{\sqrt{N}} \(\frac{M_{\rm Pl}}{m}\)^3 .
\eeq
Comparing this with eq.\eq{Qcr} suggests that, at least for not too large $N$, 
there is a range of $m$ for  forming dark dwarfs, rather than black holes.

\medskip

This is an important difference, as black holes 
lighter than $M_{\rm evap} \circa{<}M_{\rm Pl}^{5/3}/T_0^{2/3} \sim 10^{17}\,{\rm g}$ evaporate via Hawking radiation
in a cosmological time, and cannot be DM. 
Black holes with initial DM density evaporate fast enough not to damage BBN if lighter than $ 10^{10}\,{\rm g}$~\cite{2002.12778}.
Assuming that dark quarks have the DM density, black holes heavier than $M_{\rm evap}$
can only arise if $T_{\rm cr}\circa{<}3\,{\rm MeV}\, r^{1/4}$, independently of $m$,
i.e.\ above the hatched region in fig.\fig{DwarfMR}b.
To precisely predict the relative fraction of dwarfs vs black holes  we would need to know
the distribution in size of pockets, while we only computed their typical radius.
Furthermore, both dwarfs and black holes can accrete. 

On the other hand, dark dwarfs can be stable DM candidates even if lighter than $M_{\rm evap} $.
Indeed, the TOV equation implies the Buchdahl bound $R \ge \frac{9}{4} G{\cal M} > R_{\rm Sch}$, saturated for a constant density $\rho(r)$.
It implies that a finite distribution of mass, like a dark dwarf, qualitatively differs from a black hole.
Even if small compact objects emit some precursor of Hawking radiation (see~\cite{1801.03918,1802.09107}), it is negligible.
Furthermore, non-renormalizable operators that induce decays of dark quarks must be suppressed enough.
This automatically happens at larger $N$~\cite{2007.12663,2012.12087}, that also makes baryon formation more difficult.

\medskip

As an aside comment, let us assume that only black holes are formed, that later evaporate.
Even in this worst-case scenario, something interesting happened:
the physics discussed in this paper 
provides a cosmological mechanism by which the relic DM abundance can be reduced,
despite that DM particle number is conserved. Furthermore, we estimate one possible
signal: two DM dark dwarfs that collide with cross section
$\sigma v \sim (G M)^2/v$ forming a BH that evaporates into SM particles.
The resulting energy flux in SM particles received at Earth,
$dE/dt\, dS \sim \sigma v r_\odot M (\rho_{\rm DM}/M)^2\sim (M/10^{10}\,{\rm g}) \eV/{\rm Gyr\, km}^2$,
is negligibly small in the Milky Way ($r_\odot\sim 10\,{\rm kpc}$, $v \sim 10^{-3}$).

\medskip

In summary,
fig.\fig{DwarfMR} shows the minimal dark quark mass $m$ that leads to gravitational collapse as a function of $\Lambda \approx T_{\rm cr}$
and of $r$, the energy fraction in the dark sector.  
The shaded regions are excluded because $m$ is super-Planckian or because dark glue-balls over-close the Universe,
if assumed to be stable.  A wide region of parameter space is open and, as we now show, unaffected by dark baryon formation.

\subsection{Dark baryon formation}\label{DBF}
The discussion above ignored possible particle-physics processes that change the particles trapped in pockets.
While many models are possible (for example, dark quarks charged under the SM would avoid the possible extra suppression
of wall velocity), one process is possible in any model:
the dark quarks can form dark baryons and escape from the pockets, possibly preventing the formation of stable gravitational relics.
The crucial point to understand is whether the dark baryons form before of after the gravitational collapse at $R\sim R^{\rm min}_{\rm gas}$. In the former case, the dark baryons are free to escape from the pocket because they are gauge singlets. In the opposite case the gravitational energy of one dark baryon becomes bigger than its thermal energy so that it cannot escape and the system becomes gravitationally bound (unless the energy released by baryon formation is large enough to destroy the pocket).

As we now show, baryon formation has negligible effects in most of our parameter space.
Perturbative baryons are bound states with binding energies of order $E_B \sim \alpha^2_{\rm dark}m$.\footnote{For our
$\SU(N)$ gauge group, a two-body $qq$ state in the antisymmetric channel has an attractive Coulomb-like
potential $V = -(1+1/N)\alpha_{\rm dark}/2r$, not enhanced by $N$.
The large enhancement of the baryon formation cross section studied in~\cite{1811.08418}
is not present in our first order phase transition, 
as the  quark string tension vanishes for dark quarks inside the false vacuum.
}
Such bound states form
with cross section $\sigma v \sim \alpha^3_{\rm dark}/m^2$. 
The large mass of heavy quarks leads to small cross sections $\sigma$ because the binding energies $E_B$ are large.

During the phase transition, baryon formation is negligible because its time-scale 
$\tau_{\rm coll} \sim 1/\sigma n v$ is much longer than the time-scale of the phase transition
$\tau_{\rm trans} \sim  
1/g_{\rm dark}H$.
Indeed, when the pocket size reaches $R\sim R_{\rm gas}^{\rm min}$, dark quarks are not yet gravitationally bound and
have density $n_q \sim  T_{\rm cr}^3$, independently of their asymmetry $Y$.
So  $ \tau_{\rm trans}/\tau_{\rm coll} \sim   M_{\rm Pl} T_{\rm cr} \alpha^3_{\rm dark} /g_{\rm dark}m^{2} \ll 1$ in all  the parameter space relevant for us.

\smallskip

\begin{figure}[t]
$$\includegraphics[width=0.8\textwidth]{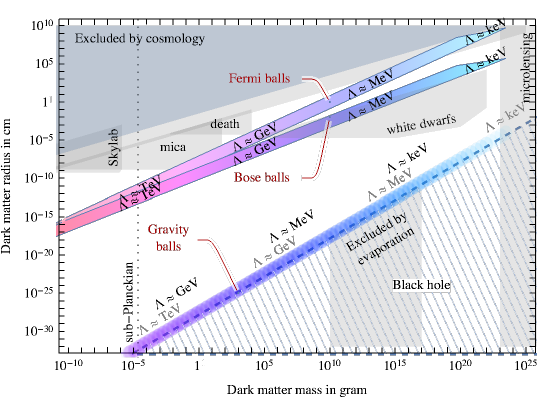}$$
\caption[Macroscopic DM]{\label{fig:MacroCompositeDM}\em 
Predictions for DM as a macroscopic object with mass $M$ and radius $R$ produced by
first-order phase transitions to confinement at scale $\Lambda$.
The upper colored  curves are the predictions for Fermi (upper) and Bose (lower) balls
in models with light dark quarks and $10^{-5}< \rho_{\rm dark}/\rho<1$.
The colored curve around the black hole boundary is the prediction for gravitational objects in models with heavy dark quarks.
Values of $\Lambda$ are indicated for $r\approx 1$ (gray) and $r\approx 10^{-10}$ (black).
Shaded regions are excluded by cosmology, Skylab, ancient mica~\cite{1410.2236}, collisions~\cite{1907.06674}, white dwarfs~\cite{1805.07381}, micro-lensing~\cite{1701.02151,1910.01285},
black hole evaporation~\cite{2002.12778}, assuming a cross section on matter $\sigma \approx \pi R^2$.
}
\end{figure}

When gravitational collapse starts at $R \sim R_{\rm gas}^{\rm min}$, the gravitational energy of one dark quark, $U_{\rm grav} \sim G_{\rm N} mM /R \sim m R_{\rm Sch}/R$, starts being bigger than its thermal energy  $\sim T_{\rm cr}$.
At this moment the energy density in dark quarks is large, $\rho_q \sim m T_{\rm cr}^3$.
During gravitational compression,  
bound state formation can release a fraction $E_B /m \sim \alpha_{\rm dark}^2$ of such large energy, and this could destroy the pockets.
This is analogous to the onset of nuclear reactions in a star, that can undergo explosive thermo-nuclear runaways.
In our case the released energy depends on density, rather than on temperature.
Indeed, after gravitational compression starts, the pocket radius $R$ decreases, 
reducing the time-scale for bound state formation, and the time-scale of the gravitational collapse:
\be
\tau_{\rm coll} \sim \frac{m^2}{\alpha_{\rm dark}^3 n_q} \sim \frac{m^2 R^3}{\alpha_{\rm dark}^3 Q},\qquad
\tau_{\rm grav} \sim \frac{M_{\rm Pl}}{\sqrt{m n}} \sim \frac{M_{\rm Pl} R^{3/2}}{\sqrt{m Q}}.
\ee
These two processes can be compared in two different ways:
\begin{itemize}
\item[1)] Bound-state formation becomes numerically significant when $\tau_{\rm coll} \sim \tau_{\rm grav} $, that
corresponds to pocket  radius
$ R^{\rm number}_{\rm ignition} \sim \sfrac{M_{\rm Pl}^{2/3} \alpha_{\rm dark}^2 Q^{1/3}}{m^{5/3}}$.
\item[2)] Since a large energy $E_B$ is released, 
bound-state formation becomes energetically significant earlier when
the total power $W_{\rm bound}  \sim E_B Q/\tau_{\rm coll}$ released by all bound-state formations is comparable to
$\dot U_{\rm grav}$.
This happens at $R^{\rm energy}_{\rm ignition} \sim M_{\rm Pl}^6 \alpha_{\rm dark}^{10}/m^7 Q$.
\end{itemize}
The released energy can possibly destroy the pockets if, at one of two ignition radii,
$E_B$ is larger than the gravitational energy of one baryon
(i.e.\ that $v_{\rm esc}< \alpha_{\rm dark}$).  In both cases 1) and 2) this happens if
\be\label{eq:bomb}
\alpha_{\rm dark} \circa{>} Q^{1/6} \left(\frac{m}{M_{\rm Pl}}\right)^{2/3}  \circa{>} \left(\frac{m}{M_{\rm Pl}}\right)^{1/6}
\ee
where, in the last step, we used the  Chandrasekhar-like threshold for gravitational collapse of eq.\eq{Qcr}.
In most of the parameter space $\alpha_{\rm dark} $, given by eq.\eq{RGEsol},
is below the critical value of eq.\eq{bomb}, such that the released energy cannot destroy the pockets.\footnote{In the presence
of a symmetric $q,\bar{q}$ component comparable to the asymmetric $q$ component, 
the gravitational collapse is modified by the energy from $q\bar{q}$ annihilations
if $\alpha_{\rm dark} \circa{>} (m/M_{\rm Pl})^{1/2}$, affecting a larger part of the parameter space. This presumably precludes the possibility of having an asymmetric number of particles or anti-particles
$N_{\rm asym} \sim \sqrt{N_{\rm sym}}$ arising accidentally inside pockets 
starting from a symmetric population with average number $N_{\rm sym}$, 
as in~\cite{2103.09827}.}

The phase where dark quarks burn into dark baryons
continues for a time-scale smaller than $\tau_{\rm coll}(R^{\rm energy}_{\rm ignition}) \sim \alpha_{\rm dark}^{27} M_{\rm Pl}^6/m^7$, certainly smaller than cosmological times. After the dark fuel is consumed, gravitational compression continues to the dark dwarf or black hole stage,
and the escape velocity reaches its final relativistic value. 

\medskip

Presumably, dark dwarfs in their final state
are in a color superconductor phase, at least after that their temperature becomes small enough~\cite{hep-ph/9711395}. 
Indeed, the system is weakly-coupled ($\alpha_{\rm dark}$ is small at $m \gg \Lambda$)
and can be approximated as a free Fermi gas up to small dark gauge interactions.
They coherently align the color of the dark quarks on the Fermi surface along the most attractive channel, which depends on the number of dark flavours. This coherent alignment can be effectively described as a condensate $\langle q q \rangle$ of Cooper pairs, which here certainly forms, 
given the absence of electromagnetic-like repulsion. 
Therefore, in this phase dark QCD would be broken by the medium, but the equation of state would be affected only marginally, given the weakness of dark-QCD interactions. In conclusion, we could safely neglect these effects in the calculation above.

\medskip

We mention one possible signal of a
dark dwarf with mass $M$ that interacts only gravitationally.
Passing through matter with speed $v\sim 10^{-3}$ it leaves two signals. 
A melting track with size $b_{\rm melt} \sim G_{\rm N}M\sqrt{m_p/m_e}/\alpha v$ due to energy losses,
and a cylindrical crack with larger size $b_{\rm break} \sim G_{\rm N}^2 M^2 m_p/\alpha v^2$ due to gravitational forces.
These are larger than the atomic size if $M \circa{>} v M_{\rm Pl}^2/\sqrt{m_e m_p}\sim 10^{14}\,{\rm g}$.
However, at this mass the flux $\Phi=\rho_{\rm DM} v/M \sim (10^{10}\,{\rm g}/M)/{\rm km}^2\,{\rm Gyr}$ is too small.
A dark dwarf crossing matter with density $\rho$
can accrete mass 
$dM/d\ell  \approx \rho\, \pi R^2 (1+v_{\rm esc}^2/v^2)$
and collapse to a black hole that evaporates via Hawking radiation
giving a visible signal even for small masses $M \sim{\rm g}$;
however the rate of this signal is again negligible small.

\section{Dark Matter as multiverse relics?}\label{DMmulti}
The idea that our universe is one anthropically selected vacuum in a wide multiverse is motivated by the unnaturalness of the vacuum energy~\cite{Weinberg:1988cp}, of the weak scale~\cite{hep-ph/9707380,1906.00986}, 
by coincidences related to light fermion masses~\cite{Carr:1979sg,Barrow:1988yia,0712.2454,0809.1647}, and possibly by 
inflation and string theory~\cite{Susskind:2003kw}. 
This speculation is compatible with our current understanding of physics, but might appear scientifically untestable:
studying the multiverse from our universe seems as hopeless as studying zoology from a zoo with one animal only.
Finding a few more animals would reduce philosophical doubts.

We explore a possibility in this sense: Dark Matter as pockets of false vacua containing their particles compressed by first order phase transitions.\footnote{Different possible cosmological
multiverse signals have been discussed in~\cite{hep-th/0505232,0704.3473} (collisions of bubbles before inflation) and~\cite{1904.00020} (bubbles of other quasi-degenerate vacua that become slightly lower in regions with high matter density).}
So far we studied this phenomenon focusing on strong gauge dynamics.
We now consider the same phenomenon in more general theories with scalars, as a multiverse with many vacua mostly comes from scalar vacuum expectation values.
As a quantum field theory example,
$2^N$ vacua can arise if each of $N$ scalars $s$ has 2 different vacuum expectation values that minimise the potential~\cite{NimaLand,1911.01441}.

\subsection{Formation of relic pockets of false vacua}
Different vacua can have different gauge groups, and thereby  different sets of vectors and light chiral fermions, plus possibly extra fermions and scalars.
We thereby consider weakly-coupled models where a DM-candidate stable particle
(a scalar or a fermion or a vector) acquires mass
$m=y s$ from the coupling $y$ (a scalar quartic or a Yukawa coupling or a gauge coupling) to a scalar $s$.
Such particle is light in a false vacuum (here set to $s\approx 0$) and heavy in the SM vacuum.

Macroscopic dark relics can form in cosmology if $s$ acquires its current vacuum expectation value $s_{\rm SM}$ 
during a first order phase transition with energy difference $\Delta V$ at temperature $T  \ll m_{\rm SM} = y s_{\rm SM}$.
The mechanism is the same discussed in gauge models: bubbles of true vacuum with $s\neq 0$ appear and expand,
but DM particles cannot cross their walls, being light in the false vacuum and heavy in the true vacuum.

\smallskip

Pockets risk being destroyed in various ways.

First, pockets risk being crunched by the kinetic energy of their walls.
This is avoided if fast enough heat flow dissipates the latent heat keeping the walls slow enough.
The strong gauge interactions models studied in this paper provide an example where this condition is over-met.
More in general, particles light only inside pockets significantly interact with their walls (as they get a mass outside),
and thereby provide a pressure that slows the walls. 
Gravitational wave signals are thereby small.

\smallskip

Furthermore, some interaction might allow trapped DM particles to become light SM particles, and thereby to get out of the pockets.
This process can be slow enough or absent.
For example DM might annihilate with $\overline{\rm DM}$ into some vector $X$ that decays back into SM fermions $f \bar f$.
If DM particles are charged under the SM gauge group, $X$ can be a light SM vector, and such process is fast.
Otherwise, $X$ could be a heavy Planck-scale vector, giving rise to rates of order $\Gamma \sim m^5/M_X^4$,
that can be as slow as proton decay. 
A similar situation arises with scalar quartics, while Yukawa couplings to heavy fermions can give larger (but still small) rates of order $\Gamma \sim m^3/M^2$.
A safer possibility is that DM carries a conserved quantum number and an asymmetry, analogously to dark baryon number in gauge models.

\smallskip

If the above two phenomena do not occur, particles which are light in a false vacuum remain trapped in false-vacuum pockets.
Such pockets are stable because their energy is less than that of free massive particles: 
$Q$ compressed particles with $m_{\rm false} \ll m_{\rm SM}$ over-compensate for the higher false-vacuum energy density.

The number of particles in a pocket with initial size $R_i$ is $Q \sim n R_i^3$, where the initial pocket radius $R_i$ can be computed 
(similarly to section~\ref{pretransition}) for any first order phase transition.
Compression stops at the radius $R$ that minimises the pocket energy, given at low temperature and in the thin-wall limit by
\beq \label{eq:Urel}
U = \frac34 \(\frac{9\pi}{4N}\)^{1/3}\frac{Q^p}{R}+ \Delta V \, \frac{4\pi R^3}{3} + \sigma \, 4\pi  R^2\eeq
where $p=4/3$, corresponding to relativistic Fermi pressure (bosons give instead $p=1$ and a different order unity pre-factor).
The term proportional to the wall energy density $\sigma$ is negligible for large $Q$.
Minimising $U$ gives the radius $R\sim (Q^{p}/\Delta V)^{1/4}$~\cite{1301.0354,1810.04360,2008.04430}.
Such pockets can be macroscopic objects with super-Planckian mass.

\subsection{Post-inflationary phase transitions}
It is usually assumed (lacking a better understanding) that a final stage of slow-roll inflation ends in the SM vacuum, 
diluting anything produced before inflation down to negligible levels.  This makes all pockets formed before inflation irrelevant.
In the multiverse context,  it is possible that slow-roll inflation ends instead in some false vacuum $X$, provided that:
\begin{itemize}
\item[1)]  its vacuum energy density
is smaller than the inflationary energy,  $V_X \circa{<}  H_{\rm infl}^2 M_{\rm Pl}^2 \sim T_{\rm RH}^4 $.
Generation of scalar fluctuations and bounds on tensor modes imply that  $V_X \circa{<} (10^{16}\GeV)^4$ is significantly sub-Planckian.
$T_{\rm RH}$ is  the maximal reheating temperature after inflation.

\item[2)] it later decays before that its vacuum energy starts extra inflation.
\end{itemize}
This would allow formation of relic pockets. Let us discuss what the two conditions imply.

\smallskip

Condition 1) means that, after slow-roll inflation, the whole landscape with vacua up to Planckian energy
is no longer accessible, since $V_X$ must be  sub-Planckian.
But it is still possible that the SM vacuum is reached after multiple phase transitions through vacua in a small fraction of the landscape.
Presumably, these accessible sub-Planckian vacua are associated to new sub-Planckian degrees of freedom in the SM vacuum.
The existence of sub-Planckian new physics is suggested by neutrino masses, 
inflation (and, possibly, gauge unification and the smallness of $\theta_{\rm QCD}$).  
From a string-theory point of view, the SM could arise from a compactification for which
some extra dimension is mildly larger than the string scale, so that the associated moduli in flux compactifications are sub-Planckian. Such moduli change some aspects or parameters of the SM and their vacuum energy is sub-Planckian too. Then, as long as the vacuum $X$ contains stable particles and these are lighter than in the SM vacuum, dark compact objects as multiverse relics could form.

\smallskip

Concerning 2), the vacuum decay rate $ \gamma \approx M^4 e^{-S}$ is estimated as follows (similar considerations hold for thermal transitions).
Assuming a quartic potential $V=V_0 - \sfrac{M^2 s^2 }{2}  - \sfrac{A s^3}{3}  + \sfrac{\lambda s^4}{4}$ gives, in the thin wall limit, a bounce action
$ S \simeq \sfrac{2048\pi^2}{3\lambda(R-1)^{3}}$, that threatens to get
 large if $\lambda$ is small and if
$R \equiv \sfrac{(V_{\rm top}-V_{\rm true})}{(V_{\rm top}-V_{\rm false})}$ is very close to 1, namely when the two vacua are almost degenerate.
This shows that fast enough vacuum decay needs strong coupling and/or
vacua which are non-degenerate enough, $R\sim{\cal O}(1)$. 
So, the same vacua with sub-Planckian energy scales $V_X$ can satisfy condition 2).

\smallskip

In summary, if some episodes of post-inflationary phase transitions were of first order type, 
surviving pockets of the decayed false vacua and of their light particles
might have remained as relics in our vacuum, and possibly be the observed Dark Matter.

\medskip

While minimality is not expected to be an ingredient in the landscape context,
we finally mention the possibility that the particles trapped in the false vacuum are SM particles, rather than new particles
that happen to be light in the false vacuum.  
This could happen if the landscape contains false vacua that differ from the SM
because of secondary aspects controlled by the vacuum expectation value of a relatively light scalar.
For example, one might have a proto-SM vacuum that differs from the SM only because light quarks
have smaller Yukawa couplings and thereby smaller masses.
In this case, the proto-QCD phase transition can have $N_f \ge 3$ light flavors
and thereby be of first order, leading to compressed pockets of proto-quarks.
Unlike in the related scenario studied in~\cite{1804.10249} there is no dangerous subsequent release of large weak-scale energy,
if the extra scalar itself does not have big vacuum energy.
As Fermi-balls with $\Lambda \approx \Lambda_{\rm QCD}$ are constrained by impacts with humans~\cite{1907.06674} (see fig.\fig{MacroCompositeDM}), physicists who worry that the lack of testability of the multiverse may kill physics, 
can now worry of being killed by multiverse signals.

\section{Conclusions}\label{concl}
We considered a dark gauge group that becomes strongly coupled at a scale $\Lambda$
in the presence of one heavy dark quark $q$ with mass $m \gg \Lambda$.
The phase transition to confinement is of first order: bubbles of the true confined vacuum appear and expand.
The large latent heat reheats the universe back up to the critical temperature $T_{\rm cr}\approx\Lambda$
keeping  the expansion of the existing bubbles slow and stopping nucleation of new bubbles.
Relic heavy quarks cannot enter the confined true-vacuum phase, unless they meet other dark quarks and form dark baryons.
When the bubbles meet and coalesce, the surviving pockets in the false unconfined vacuum
have sub-Hubble initial size estimated in section~\ref{pretransition} and keep shrinking compressing the heavy quarks in them.

\smallskip

Compression accelerates particle-physics reactions such as $q\bar q$ annihilations or (in the presence of a $q$/$\bar q$ asymmetry)
hadron formation, and pockets evaporate.
In section~\ref{fingrav} we found that an alternative final result is possible:
if dark quarks are heavy enough, pockets can gravitationally collapse under their weight before evaporating.
We found that the condition for gravitational collapse in eq.\eq{Qcr} 
is parametrically the same as the Chandrasekhar condition for black hole formation, and that,
if dark quarks have a dominant relic asymmetry,
thermo-nuclear energy from baryon formation does not stop the collapse
(under the condition in eq.\eq{bomb}). 
As a result, depending on the value of $m$, two macroscopic final states are possible:
\begin{itemize}
\item Dark dwarfs mildly below the Chandrasekhar limit.
These are acceptable DM candidates.

\item Black holes.  These are acceptable DM candidates only above the Hawking limit on evaporation.
Black holes are the only possible final state if dark quarks are bosonic, rather than fermionic.
\end{itemize}
The above dynamics depends on $m$, on $\Lambda$ and on the dark-sector temperature $T_{\rm dark}/T_{\rm SM}$.
Fig.\fig{DwarfMR}b shows the values of these parameters that lead to formation of gravitational objects with the observed DM density.
Fig.\fig{MacroCompositeDM} shows possible values of the mass $M$ and radius $R$ of our gravitational relics,
finding that they are distinct from those of non-gravitational macroscopic relics that can form in different theories with light quark flavours.
Such relics interact with SM particles via gravity and via SM gauge interactions, if dark quarks are charged under them.
Furthermore, we pointed out the relevance of a color superconductor phase.

\medskip

In the final section~\ref{DMmulti} we argued that this kind of DM candidates ---
pockets of false vacua relics compressed by a first order phase transition ---
 can arise in a multiverse context, taking into account that vacuum transitions after slow-roll inflation  
 can involve some vacua near the physical SM vacuum.
For example, DM could be pockets of compressed particles that in the false vacuum are lighter than in the SM.

\paragraph{Acknowledgements}
We thank Salvatore Bottaro, Yi-Zen Chu, Gia Dvali, Michele Redi, Gregory Ridgway and Juri Smirnov for discussions and comments.
This work was supported by the ERC grant 669668 NEO-NAT and by PRIN 2017FMJFMW.

\end{document}